\documentclass[12pt, preprint]{aastex}

\usepackage{graphicx}
\usepackage{amssymb}
\usepackage{epstopdf}
\usepackage{layout}
\usepackage{subfigure}
\slugcomment{Accepted for Publication in {\it The Astrophysical Journal}}

\shorttitle{Polarization Dependence of $\gamma\gamma$ Absorption}
\shortauthors{M. B\"ottcher}

\begin{document}

\title{The Polarization Dependence of $\gamma\gamma$ Absorption --- 
Implications for Gamma-Ray Bursts and Blazars}

\author{M. B\"ottcher\altaffilmark{1}}

\altaffiltext{1}{Centre for Space Research, North-West University, Potchefstroom,
2520, South Africa}

\begin{abstract}
This paper presents an analysis of the dependence of the opacity for high-energy
$\gamma$-rays to $\gamma\gamma$ absorption by low-energy photons, on the polarization
of the $\gamma$-ray and target photons. This process has so far only been considered
using the polarization-averaged $\gamma\gamma$ absorption cross section. It is
demonstrated that in the case of polarized $\gamma$-ray emission, subject to
source-intrinsic $\gamma\gamma$ absorption by polarized target photons, this
may lead to a slight over-estimattion of the $\gamma\gamma$ opacity by up to 
$\sim 10$~\% in the case of a perfectly ordered magnetic field. Thus, for realistic
astrophysical scenarios with partially ordered magnetic fields, the use of the
polarization-averaged $\gamma\gamma$ cross section is justified for practical
purposes, such as estimates of minimum Doppler factors inferred for $\gamma$-ray 
bursts and blazars, based on $\gamma\gamma$ transparency arguments, and this paper
quantifies the small error incurred by the unpolarized-radiation approximation. 
Furthermore, it is shown that polarization-dependent $\gamma\gamma$ absorption 
of initially polarized $\gamma$-rays can lead to a slight increase in the 
polarization beyond the spectral break caused by $\gamma\gamma$ absorption, 
to an amount distinctly different from the change in polarization expected 
if the same spectral break was produced by a break in the underlying electron 
distribution. This may serve as a diagnostic of whether $\gamma\gamma$ absorption 
is relevant in sources such as $\gamma$-ray bursts and blazars where the $\gamma$-ray 
emission may be intrinsically highly polarized. 
\end{abstract}


\section{Introduction}

It has long been recognized \citep[e.g.,][]{GS67} that high-energy $\gamma$-rays from
astronomical sources are subject to $\gamma\gamma$ absorption by low-energy target
photons, if the energy threshold for pair production $\epsilon_{\gamma} \, \epsilon_t
\, (1 - \mu) \ge 2$ is fulfilled, where $\epsilon_{\gamma}$ and $\epsilon_t$ are the
energies of the $\gamma$-ray and the soft target photon, respectively, normalized to
the electron rest-mass energy, $\epsilon = E / (m_e c^2)$, and $\mu = \cos\theta$
is the cosine of the collision angle. This process is expected to limit the very-high-energy 
(VHE) $\gamma$-ray horizon out to which VHE $\gamma$-rays may be detectable by ground-based 
$\gamma$-ray observatories due to $\gamma\gamma$ absorption on the Extragalactic Background
Light \citep[EBL; e.g.,][and references therein]{Stecker92,Finke10}, and the apparent absence of 
$\gamma\gamma$ absorption signatures also provides evidence for relativistic beaming in 
$\gamma$-ray bursts \citep[GRBs; e.g.,][]{Baring93} and blazars \citep[e.g.,][]{DG95}. 

When considering the effects of $\gamma\gamma$ absorption, to the author's knowledge, all 
previous works have used the polarization-averaged cross section for $\gamma\gamma$ absorption 
(or simplified approximations derived from it),

\begin{equation}
\sigma_{\gamma\gamma}^{\rm ave} = {3 \over 16} \, \sigma_T \, (1 - \beta^2) \left\lbrace
(3 - \beta^4) \, \ln\left( {1 + \beta \over 1 - \beta} \right) - 2 \, \beta \,
(2 - \beta^2) \right\rbrace
\label{sigma_average}
\end{equation}
\citep{JR76}, where 

\begin{equation}
\beta = \sqrt{1 - {2 \over \epsilon_{\gamma} \, \epsilon_t \, (1 - \mu)}}
\label{betacm}
\end{equation}
is the normalized (to the speed of light) velocity of the produced electron and positron 
in the center-of-momentum frame of the collision. In the case of $\gamma\gamma$ absorption
of cosmological $\gamma$-rays by the EBL, the use of the polarization-averaged cross section
is reasonable as the EBL is expected to be, on average, unpolarized on cosmological scales. 

However, this may not be the case when considering source-intrinsic $\gamma\gamma$ 
absorption in GRBs and blazars. 
In GRBs, the X-ray through $\gamma$-ray emission is commonly interpreted as synchrotron
radiation by shock-accelerated electrons, possibly with an admixture of thermal radiation
from a photosphere of the initial fireball. In an ordered magnetic field, synchrotron
radiation is expected to be polarized, and Compton scattering in the photosphere of
structured GRB outflows may also result in non-zero polarization of a possible photospheric
emission component \citep{Lundman14}. In blazars, the low-energy emission, potentially acting
as targets for intrinsic $\gamma\gamma$ absorption, is generally agreed to be produced by
synchrotron radiation of relativistic electrons \citep[see, e.g.,][for a review of blazar
emission models]{Boettcher07}. In the case of synchrotron self-Compton (SSC) radiation
or hadronic emission scenarios for the $\gamma$-ray emission, also the $\gamma$-rays are
expected to be polarized \citep{ZB13}. Therefore, when considering the intrinsic $\gamma\gamma$
opacity of high-energy $\gamma$-ray sources, the effects of polarization may not be negligible.

The study of potential effects of the polarization dependence of the $\gamma\gamma$ 
opacity in GRBs and blazars is the aim of this paper. Section \ref{crosssection} contains a
brief discussion of the polarization-dependent cross section for $\gamma\gamma$ absorption. 
Section \ref{model} presents a simple model scenario to illustrate the potential impact of
the polarization dependence of the $\gamma\gamma$ absorption cross section in the case of
intrinsic absorption of polarized $\gamma$-rays by a target photon field with the same
polarization geometry as the $\gamma$-rays, as generally expected when $\gamma$-rays and 
target photons are produced co-spatially. Section \ref{discussion} contains a discussion 
of the results.

\begin{figure}
\plotone{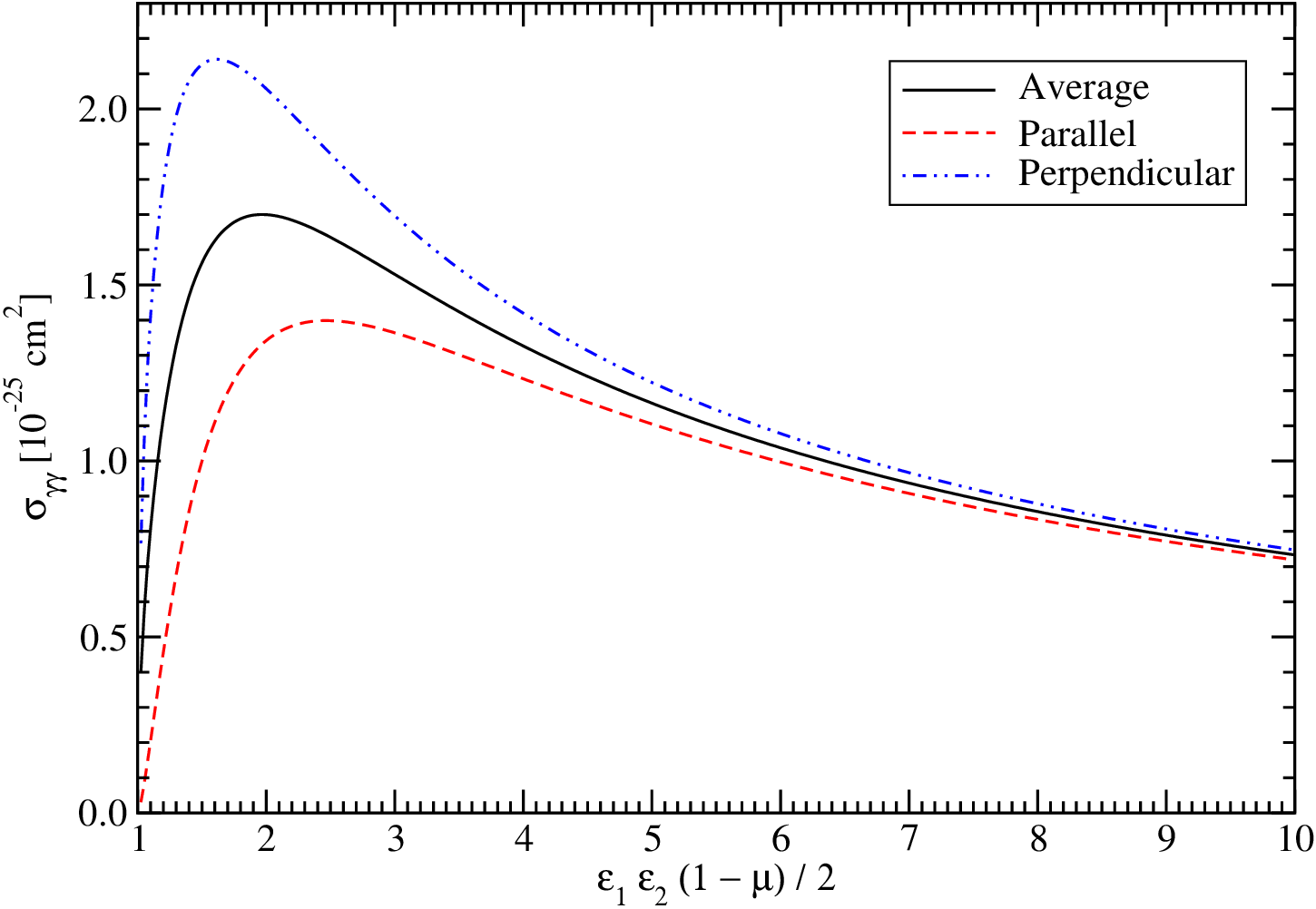}
\caption{\label{crosssection_plot}Polarization-dependent $\gamma\gamma$ absorption
cross section as a function of center-of-momentum electron energy squared, $\gamma_{\rm cm}^2
= \epsilon_{\gamma} \, \epsilon_t \, (1 - \mu) / 2$, for the case of parallel and 
perpendicular polarization directions of $\gamma$-ray and target photons, repsecitvely.}
\end{figure}

\section{\label{crosssection}The polarization-dependent $\gamma\gamma$ absorption
cross section}

The polarization-dependent cross section for $\gamma\gamma$ absorption has been calculated
by \cite{BW34}. Using the same nomenclature as employed in \cite{JR76} (e.g., Eqs. 
\ref{sigma_average} and \ref{betacm}), the cross section $\sigma_{\gamma\gamma}^{\parallel}$ 
for the case of parallel linear-polarization vectors of the $\gamma$-ray and target photon, 
is given by

\begin{equation}
\sigma_{\gamma\gamma}^{\parallel} = {3 \over 16} \, \sigma_T \, (1 - \beta^2) \left\lbrace
{5 + 2 \, \beta^2 - 3 \, \beta^4 \over 2} \, \ln\left( {1 + \beta \over 1 - \beta} \right) 
- \beta \, (5 - 3 \, \beta^2) \right\rbrace
\label{sigma_parallel}
\end{equation}

and $\sigma_{\gamma\gamma}^{\perp}$ for the case of perpendicular polarization vectors:

\begin{equation}
\sigma_{\gamma\gamma}^{\perp} = {3 \over 16} \, \sigma_T \, (1 - \beta^2) \left\lbrace
{7 - 2 \, \beta^2 - \beta^4 \over 2} \, \ln\left( {1 + \beta \over 1 - \beta} \right) 
- \beta \, (3 - \beta^2) \right\rbrace
\label{sigma_perp}
\end{equation}

One verifies easily that $(\sigma_{\gamma\gamma}^{\parallel} + \sigma_{\gamma\gamma}^{\perp})/2 
= \sigma_{\gamma\gamma}^{\rm ave}$ according to Eq. \ref{sigma_average}. These cross sections 
as a function of center-of-momentum energy are plotted in Figure \ref{crosssection_plot}. It 
can be seen that the cross section for the case of perpendicular polarization directions (a) 
peaks at lower energies than the average and the parallel case and (b) peaks at a value about 
1.5 times higher the peak cross section for the parallel case. 

This means that the $\gamma\gamma$ absorption of polarized $\gamma$-ray photons by target 
photons with identical polarization direction is suppressed compared to the unpolarized 
case (and compared to the case of absorption on target photons with polarization direction 
perpendicular to that of the $\gamma$-ray photon). This may be important in cases where
polarized $\gamma$-rays are produced co-spatially with synchrotron target photons, with
preferred electric-field vector orientations perpendicular to a globally ordered magnetic
field in the emission region. If the $\gamma$-rays are co-spatially produced, e.g., by 
SSC scattering, proton synchrotron radiation, or synchrotron emission of secondary 
particles in photo-pion induced cascade processes, then they are expected to have 
the same polarization direction as the target electron-synchrotron photons \citep{ZB13}. 
In this case, we expect two consequences of the polarization-dependence of the cross sections
(\ref{sigma_parallel}) and (\ref{sigma_perp}): 1. The overall $\gamma\gamma$ opacity is reduced
compared to $\gamma\gamma$ absorption by unpolarized photons, and 2. the degree of polarization
$\Pi$ is expected to increase in a partially self-absorbed regime \citep[see, e.g.,][for a
discussion of possible signatures of partially self-absorbed SSC emission in GRBs]{Panaitescu14}, 
because the dominant polarization direction of the $\gamma$-ray beam will be less affected by 
$\gamma\gamma$ absorption since it is primarily interacting with the sub-dominant polarization 
direction of the target photon field, and vice versa. These effects will be illustrated and 
studied more quantitatively with a simple toy model of a synchrotron-dominated $\gamma$-ray 
source in the next section.

\section{\label{model}Effects on a synchrotron-dominated $\gamma$-ray source}

In this section, we investigate the potential effects of polarization-dependent $\gamma\gamma$
absorption in an idealized case of a synchrotron-dominated $\gamma$-ray source, as commonly
assumed for the non-thermal emission from GRBs. The model assumes a spherical emission region
containing a perfectly ordered magnetic field, and considers a non-thermal electron synchrotron spectrum 
characterized by an energy index $\alpha$, i.e., an emission coefficient $j_{\nu} \propto
\nu^{-\alpha}$, extending from $\epsilon_1 = 10^{-6}$ 
without cut-off into the $\gamma$-ray regime. Additional, 
polarized high-energy and very-high-energy $\gamma$-ray photons 
may be produced, e.g., by proton synchrotron radiation or 
synchrotron emission of secondary leptons produced in photo-pion 
produced pair cascades \citep[e.g.,][]{ZB13}.
The synchrotron 
emissivity is normalized to a total synchrotron compactness $\ell = \sigma_T \, L_{\rm sy} / 
(4 \pi \, R \, \langle\epsilon\rangle m_e c^3) \sim 1$, so that $\gamma\gamma$ absorption 
effects are expected to become important at high energies. Based on the spectral index $\alpha$, 
the synchrotron spectrum is characterized by a degree of polarization 

\begin{equation}
\Pi \equiv {j_{\perp} - j_{\parallel} \over j_{\perp} + j_{\parallel}} = {\alpha + 1
\over \alpha + 5/3}
\label{Pi}
\end{equation}
from which we find the emission coefficients with electric-field vector orientations pependucular
and parallel to the magnetic field, as $j_{\perp} = j \, (1 + \Pi) / 2$ and $j_{\parallel} = j \, 
(1 - \Pi)/2$, respectively. For the case study in this section, let us consider the $\gamma\gamma$
opacity of a synchrotron $\gamma$-ray beam emitted perpendicular to the B-field. 
Due to the dependence of the synchrotron emissivity on the 
pitch angle $\chi$ of relativistic particles, $j_{\nu} (\chi)
\propto \sin^2 \chi$, the majority of polarized synchrotron 
photons is expected to be emitted in this direction ($\chi =
\pi/2$). Let us further
choose the direction of propagation of the $\gamma$-ray as the z-axis, and the B-field along the 
y-axis. We define $\theta$ as the angle that a target photon momentum makes with the z-axis (and, 
thus, the $\gamma$-ray photon momentum), and $\phi$ as the azimuthal angle around the z-axis, with
$\phi = 0$ in the (y,z) plane. The $\gamma$-ray photon then has polarization vectors given
by $\hat s_{\gamma, \perp} = \hat x$ and $\hat s_{\gamma, \parallel} = \hat y$. One then finds that 
the polarization vector $s_{t, \parallel}$ parallel to the magnetic field projection onto the plane 
perpendicular to the direction of propagation of any target photon interacting with the $\gamma$-ray 
from an angle ($\theta$, $\phi$) forms an angle $\psi$ with $s_{\gamma, \parallel}$ given by $\cos\psi = 
\hat s_{\gamma, \parallel} \cdot \hat s_{t, \parallel} = \sqrt{1 - \sin^2\theta \, \cos^2\phi}$.
Using this result, one can find the $\gamma\gamma$ absorption coefficients for $\gamma$-rays with
electric-field vectors parallel and perpendicular to the magnetic field, as

$$
\kappa_{\nu}^{\perp} = \int\limits_0^{\infty} d\epsilon \, \int\limits_{4\pi} d\Omega \, (1 - \mu)
$$
$$
\cdot \Bigl\lbrace \sigma_{\gamma\gamma}^{\parallel} \left( n^{\perp} [\epsilon, \Omega] \, \cos^2\psi
+ n^{\parallel} [\epsilon, \Omega] \, \sin^2\psi \right)
$$
\begin{equation}
+ \sigma_{\gamma\gamma}^{\perp} \left( n^{\parallel} [\epsilon, \Omega] \, \cos^2\psi
+ n^{\perp} [\epsilon, \Omega] \, \sin^2\psi \right) \Bigr\rbrace
\label{kappa_perp}
\end{equation}
and

$$
\kappa_{\nu}^{\parallel} = \int\limits_0^{\infty} d\epsilon \, \int\limits_{4\pi} d\Omega \, (1 - \mu)
$$
$$
\cdot \Bigl\lbrace \sigma_{\gamma\gamma}^{\parallel} \left( n^{\parallel} [\epsilon, \Omega] \, \cos^2\psi
+ n^{\perp} [\epsilon, \Omega] \, \sin^2\psi \right)
$$
\begin{equation}
+ \sigma_{\gamma\gamma}^{\perp} \left( n^{\perp} [\epsilon, \Omega] \, \cos^2\psi
+ n^{\parallel} [\epsilon, \Omega] \, \sin^2\psi \right) \Bigr\rbrace
\label{kappa_parallel}
\end{equation}
where the photon densities $n^{\perp}$ and $n^{\parallel}$ are evaluated based on the 
emissivities $j_{\parallel}$ and $j_{\perp}$ as outlined above.

\begin{figure}
\plotone{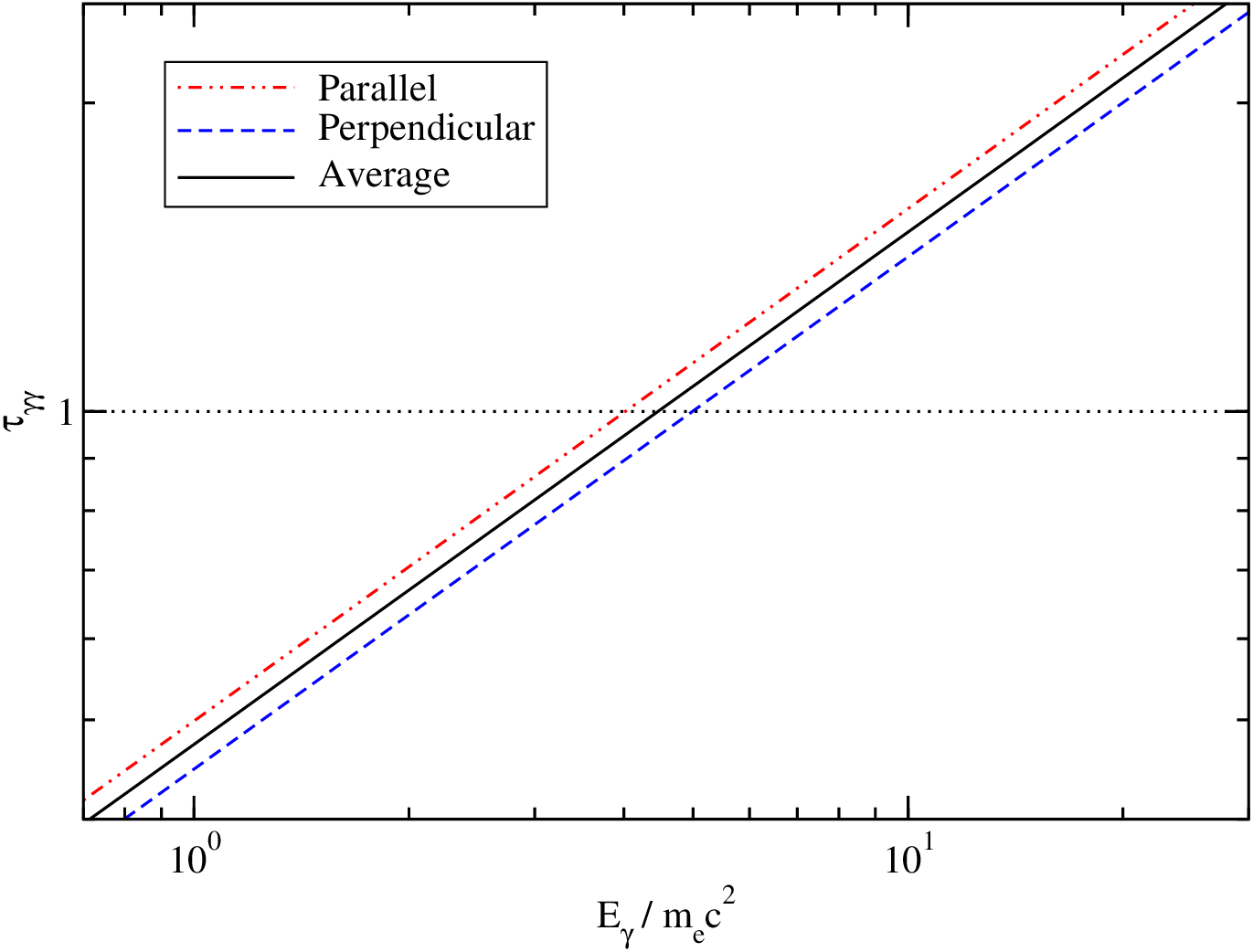}
\caption{\label{synchrotron_tau_plot}Polarization-dependent $\gamma\gamma$ opacity as a
function of $\gamma$-ray photon energy, in a synchrotron-dominated source, for photons 
propagating perpendicular to the magnetic field, in a source with compactness $\ell = 1$, 
with synchrotron spectral index $\alpha = 0.5$. The figure illustrates that the opacity
for $\gamma$-rays with electric-field vectors perpendicular to the $B$-field (i.e., the
dominant polarization direction) is about 10~\% smaller than for photons with E-field 
vectors parallel to $B$.}
\end{figure}

Figure \ref{synchrotron_tau_plot} shows the resulting $\gamma\gamma$ opacities for a source 
with compactness $\ell = 1$, and a synchrotron spectral index of $\alpha = 0.5$. Both the
polarization-dependent and the polarization-averaged opacities show the well-known energy
dependence $\tau_{\gamma\gamma} \propto \epsilon^{\alpha}$. As expected, the $\gamma\gamma$
opacity for photons with parallel polarization direction is larger than that for photons with
perpendicular (the dominant) polarization direction. However, the effect is smaller than the
difference in the peak values of the respective cross sections, since the target photon field 
is not 100~\% polarized, thus mitigating the effect. Still, the difference between the 
opacities in the two polarization directions is about 10~\% in this case. 

\begin{figure}
\plotone{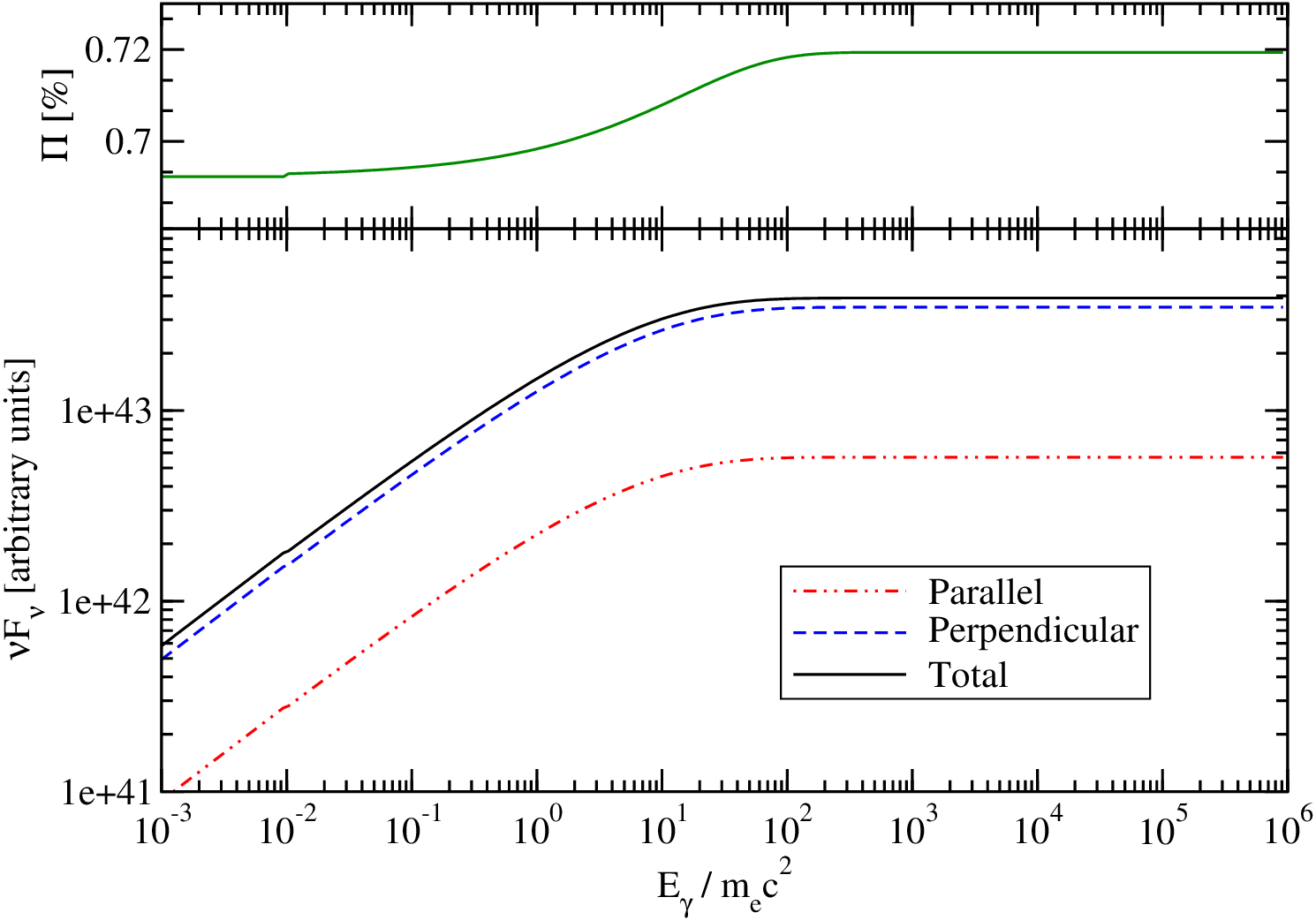}
\caption{\label{synchrotron_spectrum_plot}Polarization-dependent flux spectrum (lower panel) 
and degree of polarization as a function of photon energy (upper panel). Parameters are the
same as for Figure \ref{synchrotron_tau_plot}}
\end{figure}

Given an expected optically-thin synchrotron flux $F_{\nu}^{\rm int}$ emerging from the 
spherical synchrotron source considered here, the emerging spectrum $F_{\nu}^{\rm obs}$
including the effects of $\gamma\gamma$ absorption is calculated separately for both 
polarization directions, as

\begin{equation}
F_{\nu}^{\rm obs} = F_{\nu}^{\rm int} \, {1 - e^{-\tau_{\gamma\gamma}} \over \tau_{\gamma\gamma}}
\label{Fnu}
\end{equation}
The degree of polarization $\Pi$ of the emerging spectrum is then evaluated based on the
emerging fluxes with parallel and perpendicular polarization directions, i.e., $\Pi = 
(F_{\perp} - F_{\parallel})/(F_{\perp} + F_{\parallel})$. The result for 
our baseline example is plotted in Figure \ref{synchrotron_spectrum_plot}. One sees the
expected spectral break $\Delta\alpha_{\gamma} = \alpha_{\rm target} = \alpha = 0.5$ in 
accordance with Equation \ref{Fnu}. The top panel illustrates the effect mentioned in the
previous section, that the dominant (perpendicular) polarization direction is less affected 
by $\gamma\gamma$ absorption than the sub-dominant (parallel) one, leading to an increase
of the degree of polarization $\Pi$ towards the optically-thick regime. 

While the change of the degree of polarization, $\Delta\Pi$ expected from polarization-dependent
$\gamma\gamma$ absorption is a rather small effect ($\Delta\Pi = 2.7$~\%), it can be compared to 
the change expected if the associated spectral break is due to a break in the underlying electron 
distribution. In that case, one expects a change $\Delta\Pi = \Delta([\alpha + 1]/[\alpha + 5/3])$. 
For a break from $\alpha = 0.5 \to 1$, this would yield a change of $\Delta\Pi = 5.8$~\%. 

\begin{figure}
\plotone{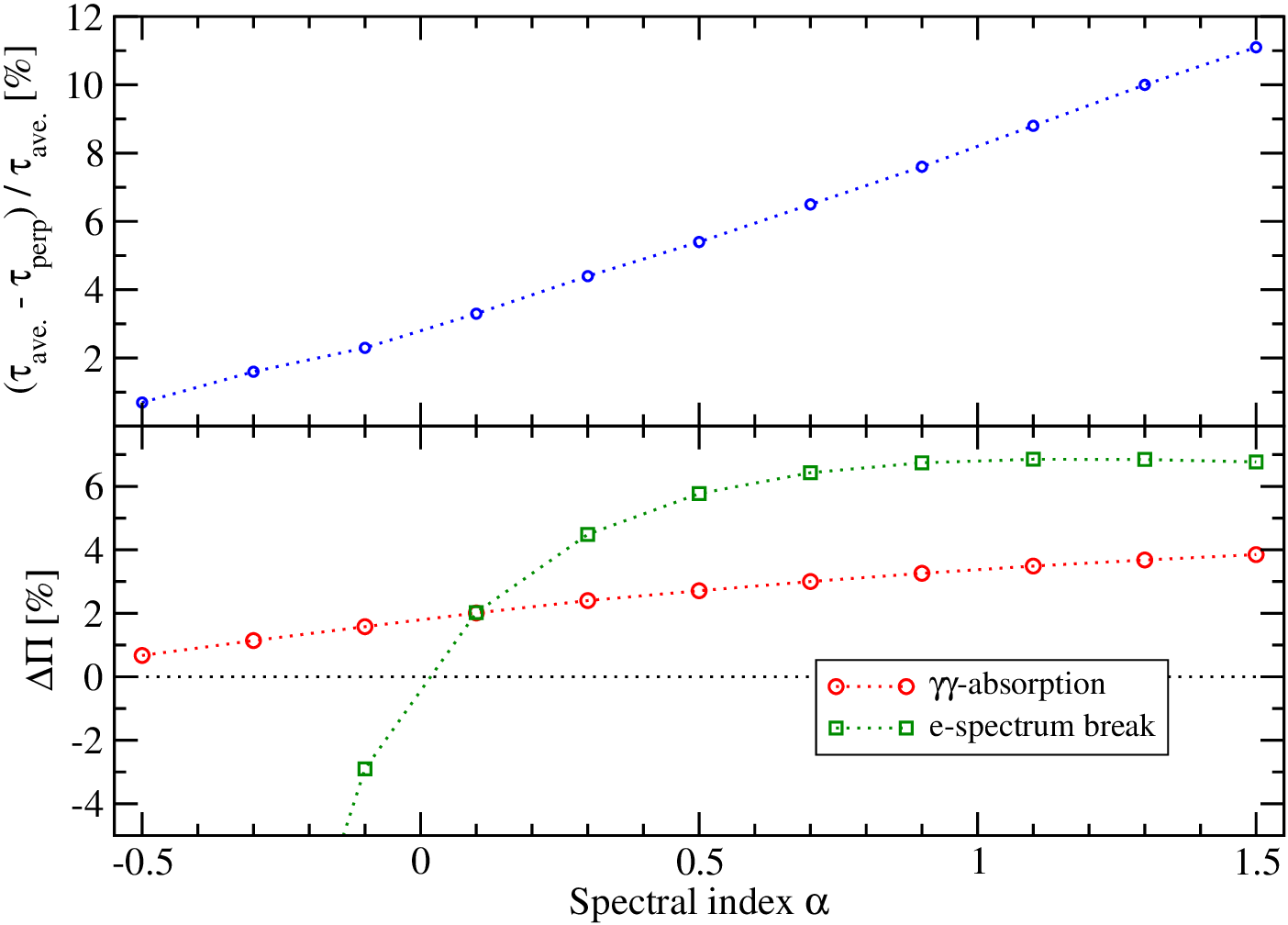}
\caption{\label{tau_alpha_plot}{\it Upper panel:} Ratio of unpolarized (i.e., average) to 
polarized (perpendicular) $\gamma\gamma$ opacity as a function of spectral index $\alpha$.
{\it Lower panel:} Change of the degree of polarization $\Pi$ across the spectral break,
as a function of spectral index $\alpha$, expected for a break caused by polarization-dependent
$\gamma\gamma$-absorption (circles) and for a break caused by a break in the underlying electron
spectrum (squares). }
\end{figure}

Figure \ref{tau_alpha_plot} illustrates how these results depend on the spectral index
$\alpha$ of the synchrotron spectrum. The upper panel shows that the use of the average
(unpolarized) $\gamma\gamma$ opacity may overestimate the actual opacity for polarized
$\gamma$-rays by up to $\sim 10$~\% for a relatively steep intrinsic spectrum. The lower
panel compares the change in $\gamma$-ray polarization $\Pi$ across the spectral break
between the two cases of the break being caused by $\gamma\gamma$ absorption and by a
break in the underlying electron distribution. While the polarization-change due to
$\gamma\gamma$ absorption is always expected to be at the $\lesssim 4$~\% level, a
break in the underlying electron distribution may cause much larger changes. Below
we will discuss whether this may be used as a diagnostic of the importance of $\gamma\gamma$
absorption in the formation of the high-energy spectra of GRBs and blazars.

\section{\label{discussion}Summary and Discussion}

The main results of the study presented in the previous sections can be summarized
as follows:

\begin{itemize}
\item The use of the unpolarized (average) $\gamma\gamma$ opacity may over-estimate
the actual $\gamma\gamma$ opacity by a small amount, up to $\sim 10$~\% in cases where
high-energy $\gamma$-rays and target photons have identical preferred polarization 
directions, and the magnetic field is perfectly ordered. This effect becomes larger 
with increasing spectral index of the target photon field.
\item Polarization-dependent $\gamma\gamma$ absorption leads to a spectral break in
the emerging $\gamma$-ray spectrum, which is accompanied by a small increase of the
percentage polarization $\Pi$, which is, for spectral indices $\alpha \gtrsim 0.2$,
smaller than the expected change in polarization resulting from a break in the underlying
electron distribution.
\end{itemize}

Both in the case of GRBs and blazars, the target photon field for $\gamma\gamma$
absorption is most likely of synchrotron origin, and therefore expected to be polarized.
The non-thermal $\gamma$-ray emission from GRBs is also commonly attributed to 
synchrotron emission (and possibly SSC radiation), while that of blazars may be
due to SSC emission or proton-induced processes, such as proton synchrotron or
synchrotron radiation from photo-pion-induced secondaries. In those cases, also
the $\gamma$-ray emission from blazars is expected to be polarized. The study 
presented here has shown that the use of the unpolarized $\gamma\gamma$ opacity
may slightly overestimate, e.g., minimum Lorentz factors of blazars and GRBs.

In our simple toy model, a perfectly ordered magnetic field has been assumed.
From the observed optical polarization, e.g., of blazars, reaching maximum values
of $\Pi \lesssim 40$~\%, one can infer that due to partial disorder in the B-field,
the degree of polarization of the $\gamma$-ray and target photon fields is reduced 
to at most about 50~\% of the level expected for a perfectly ordered B-field. 
Therefore, realistically, one may expect that $\gamma\gamma$ opacities may be
overestimated by no more than $\sim 5$~\%, which --- for most practical purposes ---
is a sufficiently small error to justify the use of the polarization-averaged
$\gamma\gamma$ crosssection. 

An additional, simplifying assumption made in our toy model
was the extension of the polarized target photon field into the
$\gamma$-ray regime without any cut-off. A high-energy cut-off 
of the synchrotron target photon spectrum would result in an
increased degree of polarization at and beyond the (normalized)
cut-off energy $\epsilon_{\rm cut}$. This will result in a 
larger effect of the polarization dependence of the $\gamma
\gamma$ absorption cross section at $\gamma$-ray photon energies 
$\epsilon_{\gamma} \lesssim 1 / \epsilon_{\rm cut}$, where the
over-estimation of the $\gamma\gamma$ opacity when using the
polarization-averaged cross section, would become more
severe than discussed above.

The measurement of high-energy polarization is a very challenging task. However,
satellite-borne instruments, such as SPI and IBIS on board the INTEGRAL satellite, 
have already been used successfully to constrain the hard X-ray / soft $\gamma$-ray 
polarization from gamma-ray bursts \citep{Dean08,Forot08}, and design studies for the 
upcoming ASTRO-H mission suggest that it may also be able to detect polarization 
in the 50 -- 200~keV energy band \citep{Tajima10}. It has also been suggested 
that the Large Area Telescope (LAT) on-board the {\it Fermi} Gamma-Ray Space 
Telescope may be able to detect $\gamma$-ray polarization in the energy range 
$\sim 30$ -- 200~MeV when considering pair-conversion events occurring in the 
Silicon layers of the detector, by taking advantage of the polarization-dependent
direction of motion of the electron-positron pairs produced in the $\gamma$-ray --
pair conversion process \citep{Buehler10}. For bright $\gamma$-ray sources, degrees 
of polarization down to $\sim 10$~\% may be detectable. However, the feasibility of
such $\gamma$-ray polarization measurements with {\it Fermi}-LAT is highly controversial.
The proposed GAMMA-LIGHT mission \citep{morselli14} may provide substantial progress
in the ability to measure $\gamma$-ray polarization, due to the absence of any tungsten
conversion layers in its design. With such advances, it may be feasible to detect the
$\gamma$-ray polarization of GRBs and blazars. This will open up the avenue to (a)
more precisely determine the expected $\gamma\gamma$ opacity constraints relevant
to these sources and (b) identify the nature of spectral breaks in the $\gamma$-ray
spectra of GRBs and blazars, which will afford deeper insight into the nature of
the underlying electron distribution and, hence, the mechanisms leading to the
acceleration of particles to ultrarelativistic energies in the relativistic jets
of GRBs and blazars.

\section*{Acknowledgments}

The author thanks Haocheng Zhang for stimulating discussions
and the anonymous referee for a helpful and constructive report
which helped improve the manuscript. He acknowledges support 
from the South African Department of Science and Technology 
through the National Research Foundation under NRF SARChI Chair 
grant No. 64789.

\end{document}